\documentclass[page-classic]{epl2} 
\title{Manifestation of three-body forces in $f_{7/2}$-shell nuclei. }

\shorttitle{Manifestation of three-body forces in $f_{7/2}$-shell nuclei.}
\author{Alexander Volya}
\shortauthor{A. Volya}

\institute{                    
Department of Physics, Florida State University, Tallahassee, FL
32306-4350, USA
}

\abstract{The traditional nuclear shell model approach is extended to include
many-body forces. The empirical Hamiltonian with a three-body force
is constructed for the identical nucleons on the $0f_{7/2}$ shell. Manifestations
of the three-body force in spectra, binding energies, seniority mixing,
particle-hole symmetry, electromagnetic and particle transition rates
are investigated. 
It is shown that in addition to the usual expansion of the valence space within the tranditional two-body shell model, the three-body component in the Hamiltonian  can be an important part improving the quality of the theoretical approach.}

\begin{document}

\maketitle
The many-body problem is central for modern physics. A path from the understanding of
interactions between fundamental constituents to that of the diverse physics
of the whole system is non-trivial and involves various entangled
routes. Among numerous issues, the questions of forces, effective
or bare, their hierarchy and renormalizations are of particular
importance. Nuclear physics is a valuable natural arena to explore
this. 

Microscopic treatments provide a remarkably accurate description of
light nuclei based on the observed, bare, two-nucleon interactions
\cite{Pieper:2001}. The same calculations indicate that the role
of three-body forces is increasingly large and non-perturbative for
heavier systems. The many-body approaches with roots in effective
interactions from mean-field (one-body) to shell model (one and two-body)
indicate a need in empirical many-body force \cite{Hees:1989,Hees:1990,Zuker:2003}.
It is established that, regardless of ab-initio interactions, work
in restricted space always gives rise to many-body forces; moreover,
renormalizations may successfully set the hierarchy of importance\cite{SCHWENK:2008,Barrett:2005}.

The goal of this work is to examine three-body forces within the nuclear
shell model (SM) approach. This includes determination of effective
interaction parameters, study of hierarchy in strength from single-particle
(s.p.) to two-body to three-body and beyond, manifestations in energy
spectra and transitions rates, comparison with different traditional
SM calculations and overall assessment for the need of beyond-two-body
SM. Previous works in this direction have shown improved description
of energy spectra \cite{Eisenstein:1973,Poves:1981,Hees:1989,Hees:1990}
and the significance of three-body monopole renormalizations \cite{Zuker:2003}. 

The effective interaction Hamiltonian is a sum $H_{k}=\sum_{n=1}^{k}H^{(n)}$
where the rotationally-invariant  $n$-body part is \begin{equation}
H^{(n)}=\sum_{\alpha\beta}\sum_{L}V_{L}^{(n)}(\alpha\beta)\sum_{M=-L}^{L}\, T_{LM}^{(n)^{\dagger}}(\alpha)\, T_{LM}^{(n)}(\beta),\label{eq:Hcn}\end{equation}
the isospin label is omitted here for the sake of simplicity. The
operators $T_{LM}^{(n)^{\dagger}}(\alpha)$ are normalized, i.e.  $\langle0|T_{L'M'}^{(n)}(\alpha')\, T_{LM}^{(n)^{\dagger}}(\alpha)|0\rangle=\delta_{\alpha\alpha'}\delta_{LL'}\delta_{MM'},$
$n$-body creation operators coupled to a total angular momentum $L$
and magnetic projection $M$, $T_{LM}^{(n)^{\dagger}}(\alpha)=\sum_{12\dots n}C_{12\dots n}^{LM}(\alpha)\, a_{1}^{\dagger}a_{2}^{\dagger}\dots a_{n}^{\dagger},$
where $1$ is the s.p. index. The traditional SM approach is limited
by the Hamiltonian $H_{2}=H^{(1)}+H^{(2)}$ which is a sum of the
s.p. ($n=1$) and the two-body ($n=2$) terms. In the two-body part the
coefficients $C_{12}^{LM}$ are proportional to the Clebsch-Gordan
coefficients and the index $\alpha$ is uniquely identified by the s.p.
levels involved. For $n>2$ the index $\alpha$ must include additional
information about the coupling scheme, the choice of which, in general,
is not unique. In numerical work it is convenient to define the coefficients
$C_{12\dots n}^{LM}(\alpha)$ and the normalized $n$-body operators
$T_{LM}^{(n)}(\alpha)$ using a set of orthogonal eigenstates $|LM\alpha\rangle=T_{LM}^{(n)^{\dagger}}(\alpha)|0\rangle$
of some $n$-particle system, see also \cite{Volya:2008}. 

Here we study identical nucleons in a single-$j$ $0f_{7/2}$ shell; see Refs.~\cite{Talmi:1957,Ginocchio:1963,McCullen:1964,Eisenstein:1973,Volya:2002PRC} for past works in this
mass region.
The best experimentally explored systems are $N=28$ isotones starting
from $^{48}$Ca considered as a core with protons filling the $0f_{7/2}$
shell and the Z=20 $^{40-48}$Ca isotopes with valence neutrons. The
experimentally known states identified with the $f_{7/2}$ valence
space are listed in Tab.~\ref{tab:Experimental-data}. %
\begin{table}
\begin{tabular}{|c|c||c|c|c||c|c|c|}
\hline 
\multicolumn{2}{|c||}{} & \multicolumn{3}{c||}{$N=28$} & \multicolumn{3}{c|}{$Z=20$}\tabularnewline
\hline 
Spin & $\nu$ & Name & Binding & 3B$f_{7/2}$ & Name & Binding & 3B$f_{7/2}$\tabularnewline
\hline
\hline 
0 & 0 & $^{48}$Ca & 0 & 0 & $^{40}$Ca & 0 & 0\tabularnewline
\hline
\hline 
7/2 & 1 & $^{49}$Sc & 9.626 & 9.753  & $^{41}$Ca & 8.360 & 8.4870 \tabularnewline
\hline
\hline 
0 & 0 & $^{50}$Ti & 21.787 & 21.713  & $^{42}$Ca & 19.843 & 19.837 \tabularnewline
\hline 
2 & 2 & 1.554 & 20.233 & 20.168  & 1.525 & 18.319 & 18.314 \tabularnewline
\hline 
4 & 2 & 2.675 & 19.112 & 19.158  & 2.752 & 17.091 & 17.172 \tabularnewline
\hline 
6 & 2 & 3.199 & 18.588 & 18.657  & 3.189 & 16.654 & 16.647 \tabularnewline
\hline
\hline 
7/2 & 1 & $^{51}$V & 29.851 & 29.954  & $^{43}$Ca & 27.776 & 27.908 \tabularnewline
\hline 
5/2 & 3 & 0.320 & 29.531 & 29.590  & 0.373 & 27.404 & 27.630\tabularnewline
\hline 
3/2 & 3 & 0.929 & 28.922 & 28.992  & 0.593 & 27.183 & 27.349\tabularnewline
\hline 
11/2 & 3 & 1.609 & 28.241 & 28.165  & 1.678 & 26.099 & 26.128 \tabularnewline
\hline 
9/2 & 3 & 1.813 & 28.037 & 28.034  & 2.094 & 25.682 & 25.747 \tabularnewline
\hline 
15/2 & 3 & 2.700 & 27.151 & 27.106  & 2.754 & 25.022 & 24.862 \tabularnewline
\hline
\hline 
0 & 0 & $^{52}$Cr & 40.355 & 40.292  & $^{44}$Ca & 38.908 & 38.736 \tabularnewline
\hline 
$2_1$ & $2^{*}$ & 1.434 & 38.921 & 38.813  & 1.157 & 37.751 & 37.509 \tabularnewline
\hline 
$4_1$ & $4^{*}$ & 2.370 & 37.986 & 38.002  & 2.283 & 36.625 & 36.570 \tabularnewline
\hline 
$4_2$ & $2^{*}$ & 2.768 & 37.587 & 37.643  & 3.044 & 35.864 & 36.009 \tabularnewline
\hline 
$2_2$ & $4^{*}$ & 2.965 & 37.390 & 37.183  & 2.657 & 36.252 & 35.741 \tabularnewline
\hline 
6 & 2 & 3.114 & 37.241 & 37.353  & 3.285 & 35.623 & 35.606 \tabularnewline
\hline 
5 & 4 & 3.616 & 36.739 & 36.789  & - & - & 35.180 \tabularnewline
\hline 
8 & 4 & 4.750 & 35.605 & 35.445  & (5.088) & (33.821) & 33.520 \tabularnewline
\hline
\hline 
7/2 & 1 & $^{53}$Mn & 46.915 & 47.009  & $^{45}$Ca & 46.323 & 46.406\tabularnewline
\hline 
5/2 & 3 & 0.378 & 46.537 & 46.560  & 0.174 & 46.149 & 46.280 \tabularnewline
\hline 
3/2 & 3 & 1.290 & 45.625 & 45.695  & 1.435 & 44.888 & 44.991 \tabularnewline
\hline 
11/2 & 3 & 1.441 & 45.474 & 45.454  & 1.554 & 44.769 & 44.763 \tabularnewline
\hline 
9/2 & 3 & 1.620 & 45.295 & 45.309  & - & - & 44.933 \tabularnewline
\hline 
15/2 & 3 & 2.693 & 44.222 & 44.175  & (2.878) & (43.445) & 43.214 \tabularnewline
\hline
\hline 
0 & 0 & $^{54}$Fe & 55.769 & 55.712  & $^{46}$Ca & 56.717 & 56.728, \tabularnewline
\hline 
2 & 2 & 1.408 & 54.360 & 54.286  & 1.346 & 55.371 & 55.501\tabularnewline
\hline 
4 & 2 & 2.538 & 53.230 & 53.307  & 2.575 & 54.142 & 54.332 \tabularnewline
\hline 
6 & 2 & 2.949 & 52.819 & 52.890  & 2.974 & 53.743 & 53.659 \tabularnewline
\hline
\hline 
7/2 & 2 & $^{55}$Co & 60.833 & 60.893  & $^{47}$Ca & 63.993 & 64.014 \tabularnewline
\hline
\hline 
0 & 0 & $^{56}$Ni & 67.998 & 67.950 & $^{48}$Ca & 73.938 & 73.846\tabularnewline
\hline
\end{tabular}\caption{States in $f_{7/2}$ valence space with spin and seniority listed
in the first and second columns. A $^{*}$ denotes seniority mixed
states in 3B$f_{7/2}$ and the seniority shown reflects the seniority assignment in the 2B$f_{7/2}$ model. In order to distinguish among the mixed states of the same spin-label (first column) we include an additional subscript reflecting the order in which they appear in the energy spectrum. Following are columns with data for $N=28$
isotones and $Z=20$ isotopes. Three columns for each type of valence
particles list name and excitation energy, experimental binding energy,
and energy from the three-body SM calculation discussed in the text.
All energies are in MeV's. \label{tab:Experimental-data}}

\end{table}

The three-body interactions influence nuclear masses and result in
important monopole terms \cite{Zuker:2003}. The violation of particle-hole
symmetry is related to this. Within the traditional SM in single $j$-shell, with occupancy $\Omega=2j+1,$
this symmetry makes the spectra of the $N$ and $\tilde{N}$-particle
systems ($\tilde{N}=\Omega-N$) identical, apart from a constant shift in energy.  Indeed, the
particle-hole conjugation $\mathcal{C}$ defined with $\tilde{a}_{jm}^{\dagger}\equiv\mathcal{C}a_{jm}^{\dagger}\mathcal{C}^{-1}=(-1)^{j-m}a_{j-m}$
transforms an arbitrary $n$-body interaction into itself plus some
Hamiltonian of a lower interaction-rank $H'_{n-1},$ as follows: $\tilde{H}^{(n)}=(-1)^{n}H^{(n)}+H'_{n-1}.$
The $n=1$ case corresponds to the particle number $\tilde{N}=-N+\Omega.$
For the $n=2$ we obtain a monopole shift \begin{equation}
\tilde{H}^{(2)}=H^{(2)}+(\Omega-2N)M,\,\, M=\frac{1}{\Omega}\sum(2L+1)V_{L}^{(2)}.\label{eq:M}\end{equation}
Any $H^{(1)}$ is proportional to $N$ and is thus a constant of motion,
which explains the particle-hole symmetry for the two-body Hamiltonian.
The $n\ge3$ interactions violate this symmetry leading to different
excitation spectra of $N$- and $\tilde{N}$-particle systems.
The deviations from exact particle-hole symmetry are seen in the experimental
data, Tab.~\ref{tab:Experimental-data}. The excitation energies of
$\nu=2$ states in $N=2$ system are systematically higher than those
in the 6-particle case pointing on a reduced ground state binding. 

The $j=7/2$ is the largest single-$j$ shell for which the seniority, the number
$\nu$ of unpaired nucleons, is an integral of motion
for any one- and two-body interaction \cite{Schwartz:1954,Talmi:1957}.
It is established experimentally that seniorities are mixed \cite{Armstrong:1965,Bjerregaard:1967}.
Configurations beyond the single-$j$ shell \cite{Engeland:1966,Auerbach:1967,Poves:1981,Poves:2001}
have been suggested to explain the effects; however, the possible presence
of the three-body force must be addressed. In a single-$j$ shell the pair
operators $T_{00}^{(2)}$, $T_{00}^{(2)\dagger}$ and the particle
number $N$ form an SU(2) rotational quasispin group. The quantum
numbers $\nu$ and $N$ are associated with this group. The invariance
under quasispin rotations relates states of the same $\nu$ but different
particle numbers $N$. For example, excitation energies of $\nu=2$
states are identical in all even-particle systems. The classification of operators according to quasispin leads to selection rules. 
The
s.p. operators associated with the particle transfer permit seniority
change $\Delta\nu=1$. The reactions $^{51}$V($^{3}$He, d)$^{52}$Cr
and $^{43}$Ca(d,p)$^{44}$Ca show seniority mixing as the $\nu=4$ final
states are populated \cite{Auerbach:1967,Bjerregaard:1967}. The
one-body multipole operators are quasispin scalars for odd angular
momentum, and quasispin vectors for even. Thus, the $M1$ electromagnetic
transitions do not change quasispin. In the mid-shell the quasi-vector
$E2$ transitions between states of the same seniority are forbidden.
The seniority mixing between the $\nu=2$ and $\nu=4$ pairs of $2^{+}$
and $4^{+}$states is expected in the mid-shell nuclei $^{52}$Cr
and $^{44}$Ca. Seniority can be used to classify the many-body operators
$T_{LM}^{(n)}$ and the interaction parameters. The three-body interactions
mix seniorities with the exception of interaction between $\nu=1$
nucleon triplets given by the strength $V_{7/2}^{(3)}.$ 

To determine the interaction parameters of $H_{3}$ we conduct a full
least-square fit to the data points in Tab.~\ref{tab:Experimental-data}.
This empirical method, which dates back to Refs.~\cite{Bacher:1934,Talmi:1957},
is a part of the most successful SM techniques today \cite{Brown:2006}.
Our procedure is similar to a two-body fit in sec. 3.2 of Ref.~\cite{Lawson:1980},
but here the fit is nonlinear and requires iterations due to seniority
mixing. In Tab.~\ref{tab:The-parameters} the resulting parameters
are listed for the proton (fixed $N=28$) system and neutron (fixed $Z=20$) system.
The two columns in each case correspond to fits without (2B$f_{7/2}$
left) and with (3B$f_{7/2}$ right) the three-body forces. The root-mean-square (RMS)
deviation  is given for each fit. The confidence limits can be
inferred from variances for each fit parameter given within the parentheses. 

\begin{table}
\begin{tabular}{|c|c|c||c|c|}
\hline 
 & \multicolumn{2}{c||}{$N$=28} & \multicolumn{2}{c|}{$Z$=20}\tabularnewline
\hline
\hline 
 & 2B$f_{7/2}$ & 3B$f_{7/2}$ & 2B$f_{7/2}$ & 3B$f_{7/2}$\tabularnewline
\hline
\hline 
$\epsilon$ & -9827(16)  & -9753(30) & -8542(35)  & -8487(72) \tabularnewline
\hline
\hline 
$V_{0}^{(2)}$ & -2033(60)  & -2207(97)  & -2727(122) & -2863(229) \tabularnewline
\hline 
$V_{2}^{(2)}$ & -587(39)  & -661(72)  & -1347(87)  & -1340(176) \tabularnewline
\hline 
$V_{4}^{(2)}$ & 443(25)  & 348(50) & -164(49)  & -198(130) \tabularnewline
\hline 
$V_{6}^{(2)}$ & 887(20) & 849(38) & 411(43) & 327(98) \tabularnewline
\hline
\hline 
$V_{7/2}^{(3)}$ &  & 55(28)  &  & 53(70) \tabularnewline
\hline 
$V_{5/2}^{(3)}$ &  & -18(70)  &  & 2(185) \tabularnewline
\hline 
$V_{3/2}^{(3)}$ &  & -128(88)  &  & -559(273) \tabularnewline
\hline 
$V_{11/2}^{(3)}$ &  & 102(43)  &  &  51(130) \tabularnewline
\hline 
$V_{9/2}^{(3)}$ &  & 122(41) &  & 272(98) \tabularnewline
\hline 
$V_{15/2}^{(3)}$ &  & -53(29) &  & -24(73)\tabularnewline
\hline
\hline 
RMS & 120 & 80 & 220 & 170\tabularnewline
\hline
\end{tabular}\caption{Interaction parameters of 2B$f_{7/2}$ and 3B$f_{7/2}$ SM Hamiltonians
determined with the least-square fit are given in keV's.\label{tab:The-parameters} }

\end{table}

The lowering of the RMS deviation is the first evidence in support
of the three-body forces; for $Z=28$ isotones it drops from 120keV
to about 80keV. All three-body parameters appear to be equally important,
excluding any one of them from the fit raises the RMS by about 10\%. In contrast, inclusion
of a four-body monopole force based on $\nu=0,$ $L=0$ operator led
to no improvement. The fit parameters remain stable within quoted
error-bars if some questionable data-points are removed. The energies
resulting from the three-body fit are listed in Tab.~\ref{tab:Experimental-data}.
Based on the RMS alone this description of data is quite good, even in comparison with the available large-scale two-body SM calculations
in the expanded model space \cite{Lips:1970,Poves:2001}. 

The proton and neutron effective Hamiltonians are different, 
Tab.~\ref{tab:The-parameters}. The s.p. energies reflect different mean
fields; and the two-body parameters, especially for higher $L,$ highlight
the contribution from the long range Coulomb force. However, within
the error-bars the three-body part of the Hamiltonians appears to
be the same which relates these terms to isospin-invariant strong force.

A skeptic may question some experimental states included in the fit,
thus we conduct a minimal fit considering binding energies of ground
states only, similar to Ref.~\cite{Talmi:1957}. We include a
seniority conserving part given by $V_{7/2}^{(3)}$ with $\nu=1$
triplet operator $T_{jm}^{(3)}\sim a_{jm}^{\dagger}T_{00}^{(2)}$.
This interaction is the main three-body contribution to binding and
is equivalent to a density-dependent pairing force \cite{Zelevinsky:2006}.
In a single-$j$ model it can be treated exactly with a renormalized
particle-number-dependent pairing strength $V_{0}^{(2)'}=V_{0}^{(2)}+\Omega\frac{N-2}{\Omega-2}V_{j}^{(3)}.$
From relations in Refs.~\cite{Talmi:1957,Volya:2002PRC}, the ground
state energies with $\nu=0$ or 1 are

\begin{equation}
E=\epsilon N+\frac{N-\nu}{\Omega-2}\left((\Omega-N-\nu)\frac{V_{0}^{(2)'}}{2}+(N-2+\nu)M'\right),\label{eq:Egs}\end{equation}
where $M'=M+\frac{N-2}{\Omega-2}V_j^{(3)}.$
With a linear least-square fit and Eq.~(\ref{eq:Egs}) we determine
s.p. energy $\epsilon,$ pairing $V_{0}^{(2)},$ monopole $M$, and
3-body interaction $V_{7/2}^{(3)}$ (\ref{eq:Egs}) using 8 binding
energies. The results, shown in Tab~\ref{tab:BE}, are consistent
with the full fit in Tab.~\ref{tab:The-parameters}, the repulsive
nature of the monopole $V_{7/2}^{(3)}$ is in agreement with other
works \cite{Coon:1995}. 

\begin{table}
\begin{tabular}{|c|c|c||c|c|}
\hline 
 & \multicolumn{2}{c||}{$N=28$} & \multicolumn{2}{c|}{$Z=20$}\tabularnewline
\hline 
$\epsilon$ & -9703(40) & -9692(40) & -8423(51) & -8403(55)\tabularnewline
\hline 
$V_{0}^{(2)}$ & -2354(80) & -2409(110) & -3006(120) & -3105(156)\tabularnewline
\hline 
$M$ & 1196(40) & 1166(50) & -823(55) & -876(76)\tabularnewline
\hline 
$V_{7/2}^{(3)}$ & - & 18(20) & - &  31(31)\tabularnewline
\hline
RMS & 50 & 46 & 73 & 65\tabularnewline
\hline
\end{tabular}\caption{Interaction parameters for the minimal $f_{7/2}$ SM determined with
the linear least-squared fit of 8 binding energies. The
variances for each parameter are shown within parentheses. The two columns for isotopes
and isotones are fits without and with the three-body term. \label{tab:BE}}

\end{table}

\begin{figure}
\includegraphics[width=2.5in]{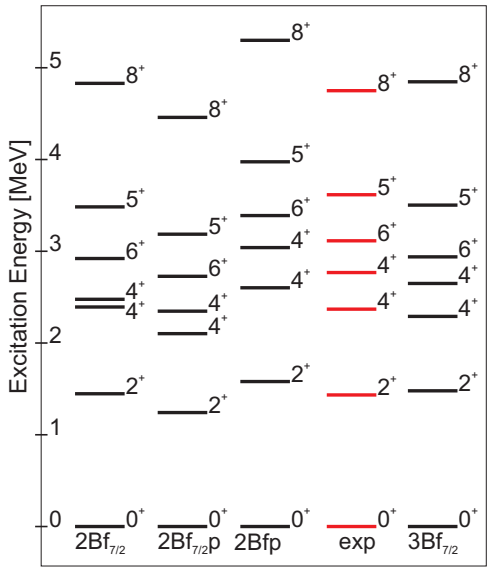}

\caption{Spectrum of $^{52}$Cr.\label{fig:Spectrum-of-52Cr}}

\end{figure}

Next we concentrate on $^{52}$Cr, Fig.~\ref{fig:Spectrum-of-52Cr}.
In addition to the 2B$f_{7/2}$ and 3B$f_{7/2}$ interactions from 
Tab.~\ref{tab:The-parameters} we perform a large-scale SM calculations
2B$f_{7/2}p$ (which includes $p_{1/2}$ and $p_{3/2}$) and 2B$fp$ (entire
$fp$-shell, truncated to $10^{7}$ projected m-scheme states) using
FPBP two-body SM Hamiltonian \cite{Brown:2001}. The 2B$f_{7/2}p$
model and its results are very close to the more restricted SM calculations
in Ref.~\cite{Lips:1970}. %

The seniority mixing between the neighboring $4_{1}^{+}$ and $4_{2}^{+}$
states leads to level repulsion. The observed energy difference of
400 keV is not reproduced by 2B$f_{7/2}$ (84 keV). The discrepancy
remains in some of the extended two-body models: 2B$f_{7/2}p$ and \cite{Lips:1970}
(200 keV). The full 2B$fp$ replicates the splitting, but possibly at the expense
of excessive intruder admixtures which distort the spectrum, see also Ref.~\cite{Speidel:2000}. The 3B$f_{7/2}$
is the best in reproducing the experimental spectrum, Fig.~\ref{fig:Spectrum-of-52Cr}.
The 3B$f_{7/2}$ model predicts seniority mixing; the $\nu(4_{1}^{+})=2.82$
and $\nu(4_{2}^{+})=2.71$ are inferred from the expectation value
of the pair operator $\langle T_{00}^{(2)^{\dagger}}T_{00}^{(2)}\rangle=(N-\nu)(2j+3-N-\nu)/(4j+2).$
The $2_{1}^{+}$, however, is relatively pure with $\nu(2_{1}^{+})=2.006$. 

Violations of the quasispin selection rules are observed in nuclei \cite{Armstrong:1965,Bjerregaard:1967,McCullen:1964,Monahan:1968,Pellegrini:1973}.
The two-body shell models beyond single-$j$ $f_{7/2}$ are used to explain this seniority mixing \cite{Auerbach:1967,Engeland:1966,Ginocchio:1963,Lips:1970,McCullen:1964}, however there are inconsistencies, suggesting a more in-depth consideration. 
Large variations of effective charges
are needed to explain
electromagnetic transitions \cite{Brown:1974}. The particle transfer spectroscopic
factors show excessive amount of components outside the $f_{7/2}$
valence space \cite{Armstrong:1965}. 
Furthermore, while going beyond the single $j$-shell is imperative to account for the observed $g$ factors \cite{Lawson:1980}, the agreement between the experiment and the common large-scale shell model results is not perfect, Ref.~\cite{Speidel:2000}, which may be a  hint for the presence of a three-body force.

In Tab.~\ref{tab:B(E2)} B$(E2)$
transitions rates from the four models are compared to experiment. The
combination of nuclear radial overlap and effective charge is normalized
using the observed $E2$ rate for the transition $2_{1}\rightarrow0$
in the 2B$f_{7/2}$, 2B$f_{7/2}p,$ and 2B$fp$ models. The normalization parameter
for the 3B$f_{7/2}$ model is identical to the one used in 2B$f_{7/2}.$
The insignificant difference in $2_{1}\rightarrow0$ B$(E2)$
between the 3B$f_{7/2}$ and 2B$f_{7/2}$ shows a small admixture of $\nu=4$
in the $2_{1}^{+}$ state. The $\nu=4$ and $\nu=2$ mixing in $4_{1}^{+}$
and $4_{2}^{+}$ effects the transitions involving these states. For example,
$E2$ transitions $4_{2}\rightarrow2_{1}$ and $6\rightarrow4_{2}$
are no longer forbidden. 

The last row in Tab.~\ref{tab:B(E2)} compares the quadrupole moment of the first $2^+$ state in $^{52}$Cr between models and experiment. It is quoted in units of $e$ fm$^{2}.$ Because of the seniority selection rule the quadrupole moment is exactly zero in the 2B$f_{7/2}$ model; the small seniority mixing between $2^{+}$ states in the 3B$f_{7/2}$ model leads to the quadrupole moment of a correct sign but it is lower than the observed one. The results from extended two-body models in  Tab.~\ref{tab:B(E2)} and in the literature \cite{Mukherjee:1984,Ernst:2000,Honma:2004} improve the picture. However, strong configuration mixing is needed to generate seniority mixing which is inconsistent with $g$ factor measurements \cite{Speidel:2000} and results in a quadrupole moment higher than observed. This again invites the possibility of an additional three-body Hamiltonian.

\begin{table}
\begin{tabular}{|c|c|c|c|c|c|}
\hline 
 & 2B$f_{7/2}$ & 2B$f_{7/2}p$ & 2B$fp$ & 3B$f_{7/2}$ & Experiment\tabularnewline
\hline
\hline 
$2_{1}\rightarrow0$$^{(*)}$ & 118.0 & 118.0 & 118 & 117.5 & 118$\pm35$\tabularnewline
\hline 
$4_{1}\rightarrow2_{1}$ & 130.4 & 122.5 & 105.8 & 73.2 & 83$\pm15$$^{(1,2)}$\tabularnewline
\hline 
$4_{2}\rightarrow2_{1}$ & 0 & 3.3 & 15.1 & 56.8 & 69$\pm18$\tabularnewline
\hline 
$4_{2}\rightarrow4_{1}$ & 125.2 & 59.3 & 2.6 & 0.5 & \tabularnewline
\hline 
$2_{2}\rightarrow0$ & 0 & 0.003 & 0.9 & 0.5 & 0.06$\pm0.05$\tabularnewline
\hline 
$2_{2}\rightarrow2_{1}$ & 119.2 & 102.2 & 101.9 & 117.1 & 150$\pm35$\tabularnewline
\hline 
$2_{2}\rightarrow4_{1}$ & 0 & 10.8 & 34.4 & 19.9 & \tabularnewline
\hline 
$2_{2}\rightarrow4_{2}$ & 57.8 & 7.2 & 5.2 & 38.7 & \tabularnewline
\hline 
$6\rightarrow4_{1}$ & 108.9 & 86.2 & 56.3 & 57.8 & 59$\pm20^{(1)}$\tabularnewline
\hline
$6\rightarrow4_{2}$ & 0 & 9.3 & 27.6 & 51.1 & 30$\pm10^{(1)}$\tabularnewline
\hline
\hline
$Q(2_1^+)\,\,\,e\,$fm$^2$ & 0 &-13.0 & -13.4$^{(3)}$ & -2.4 & $-8.2 \pm 1.6^{(4)}$\tabularnewline
\hline
\end{tabular}

\caption{B$(E2)$ transition summary on $^{52}$Cr expressed in units $e^{2}\,$fm$^{4}$, the last row being the quadrupole moment of the first $2^+$ state in units
$e\,$fm$^{2}$.
The data is taken from \cite{Huo:2007}.
The zeros in the second column for the 2B$f_{7/2}$ model are the results of the mid-shell seniority selection rules. In this model the states $2_1,\,2_2,\,4_1,$ and $4_2$ have seniorities 2, 4, 4, and 2, respectively, see 
Tab.~\ref{tab:Experimental-data}. 
$^{(*)}$In 2B $f_{7/2}p$
and 2B$fp$ models we use $0.5$ (neutron) and $1.5$ (proton) effective
charges. The overall radial scaling is fixed by B(E2,$2_{1}\rightarrow 0$).
$^{(1)}$The life-time error-bars are used. $^{(2)}$There are conflicting
experimental results on the life-time; here we use DSAM (HI, x$n\gamma$) data from Ref.~\cite{Huo:2007}, which is consistent with \cite{Brown:1974}.
$^{(3)}$ Other independent large-scale shell model calculations obtain similar values for the quadrupole moment for the $2_1^+$ state: -12.3  \cite{Honma:2004}, -15.0, -16.2 and -17.5 \cite{Mukherjee:1984}, all in units $e\,$fm$^{2}.$ $^{(4)}$ The experimental data is from \cite{Raghavan:1989}.
\label{tab:B(E2)}}

\end{table}

In Tab.~\ref{tab:Proton-removal-spectrascopic} proton removal spectroscopic
factors are compared between theoretical models and experiment $^{51}$V($^{3}$He,d)$^{52}$Cr
\cite{Armstrong:1965}. The 3B$f_{7/2}$ model is good
in its description of observation especially for the $4_{1}^{+}$
and $4_{2}^{+}$ states. It was pointed out in Ref.~\cite{Armstrong:1965}
that the spectroscopic factors for the $4^{+}$ states probe the $\nu=2$ component,
and, thus, their sum within the $f_{7/2}$ valence space is 4/3; this result
is consistent with the observation \cite{Armstrong:1965} but does not support
too much configuration mixing from outside the $f_{7/2}$ shell which would reduce the spectroscopic factors and their sum. 

\begin{table}
\begin{tabular}{|c|c|c|c|c|c|}
\hline 
 & 2B$f_{7/2}$ & 2B$f_{7/2}p$ & 2B$fp$  & 3B$f_{7/2}$ & Exp\tabularnewline
\hline
\hline 
$0^{+}$ & 4.00 & 3.73 & 3.40 & 4.00 & 4.00\tabularnewline
\hline 
$2{}_{1}^{+}$ & 1.33 & 1.14 & 0.94 & 1.33 & 1.08\tabularnewline
\hline 
$4{}_{1}^{+}$ & 0.00 & 0.13 & 0.34 & 0.63 & 0.51\tabularnewline
\hline 
$4{}_{2}^{+}$ & 1.33 & 1.11 & 0.70 & 0.71 & 0.81\tabularnewline
\hline 
$6^{+}$ & 1.33 & 1.28 & 1.28 & 1.33 & 1.31\tabularnewline
\hline
\end{tabular}

\caption{Proton removal spectroscopic factors. The experimental data is taken
from $^{51}$V($^{3}$He,d)$^{52}$Cr reaction \cite{Armstrong:1965}.
Within error-bars of about 0.1 this data is consistent with other
results \cite{Huo:2007}. \label{tab:Proton-removal-spectrascopic}}

\end{table}

To conclude, the study of nuclei in $0f_{7/2}$ shell shows evidence
for three-body forces. We extend the traditional shell model approach
by including three-body forces into consideration, we find that a
successful set of interaction parameters can be determined with an
empirical fitting procedure. With a few new parameters a sizable improvement
in the description of experimental data is obtained. The apparent hierarchy
of contributions from one-body mean-field, to two-body, to three-body
and beyond is significant; it assures the possibility of high precision
configuration-interaction methods in restricted space and supports
ideas about renormalization of interactions. The three-body forces
observed in this study appear to be isospin invariant. The new Hamiltonian
with three-body forces describes well the observables that are sensitive to such forces and to seniority, and particle-hole symmetries that are violated in their presence.  A good experimental agreement is obtained for the features of spectra, for electromagnetic
E2 transition rates, and for spectroscopic factors. While good agreement with experiment can also be obtained by configuration mixing within some advanced large-scale two-body shell model calculations, it appears that the three-body effective interaction has a somewhat different manifestation which can be experimentally identified. In particular, with the three-body force, the seniority forbidden transitions become allowed without generating excessively large quadrupole moment or, in the case of spectroscopic factors, without loosing too much strength away from a single-$j$ shell.  
Unlike the fit to experimental data discussed in this work, it appears to be difficult to fit the large-scale shell model results with a single-$j$ Hamiltonian containing a three-body force. This again suggests that configuration mixing in not necessarily equivalent to a three-body force. 
The work in this direction is to be continued.
It is important to conduct similar phenomenological investigations for other mass regions
and model spaces; on the other side, renormalization techniques that
would link fundamental and phenomenological forces \cite{SCHWENK:2008}
have to be searched for. 

\acknowledgments
The author is thankful to N. Auerbach and V. Zelevinsky for enlightening
discussions. Support from the U. S. Department of Energy, grant DE-FG02-92ER40750
is acknowledged. 
%\bibliographystyle{eplbib}
%\bibliography{n-body}

\end{document}